\documentclass[twocolumn]{aastex62}
\usepackage[mediumspace,mediumqspace,Grey,squaren]{SIunits}

\graphicspath{{./}{figures/}}

\submitjournal{ApJ}

\shorttitle{Stellar populations in Lindsay-113}
\shortauthors{Li et al.}

\begin{document}

\title{When does the onset of multiple stellar populations in star
 clusters occur-II: No evidence of multiple stellar populations in 
 Lindsay 113}

\correspondingauthor{Chengyuan Li}
\email{lichengy5@mail.sysu.edu.cn}

\author{Chengyuan Li} 
\affil{School of Physics and Astronomy, Sun Yat-sen University, Zhuhai 519082, China}
\affiliation{Key Laboratory for Optical Astronomy, National
Astronomical Observatories, Chinese Academy of Sciences, 20A Datun
Road, Beijing 100101, China}

\author{Yue Wang} 
\affiliation{Key Laboratory for Optical Astronomy, National
Astronomical Observatories, Chinese Academy of Sciences, 20A Datun
Road, Beijing 100101, China}

\author{Antonino P. Milone}
\affiliation{Dipartimento di Fisica e Astronomia ``Galileo Galilei'', Univ. di Padova, Vicolo dell'Osservatorio 3, Padova, IT-35122}

\begin{abstract}
The presence of multiple populations (MPs) in almost all globular clusters (GCs) 
older than $\sim$10 Gyr, has caught lots of attention. Recently, cumulative 
evidence indicates that extragalactic GCs that are older than 2 Gyr, seem to 
also harbor MPs, however, those that are younger than 2 Gyr do not. These 
observations seem to imply that age is a primary property that  
controls the presence of MPs in star clusters. However, because of the lack 
of studies of {intermediate-age ($\sim$2--6 Gyr-old),} low mass clusters, it is unclear if 
the cluster mass, in addition to age, also plays a role in the occurrence of MPs. 
In this work, we studied a $\sim$4 Gyr-old, low mass ($\sim$23,000 $M_{\odot}$) 
cluster, Lindsay-113, in the Small Magellanic Cloud. Using {\sl Hubble Space Telescope} 
photometry, we find that the width of the red-giant branch in this cluster, when measured 
in a specific color index that is sensitive to star-to-star chemical variations, {can 
be adequately explained by a ``simple'' stellar population model with {some possible} 
noises contributed from measurement uncertainty, photometric artifact, as well as  
differential reddening.} The comparison of observations with predictions from 
synthetic spectra indicates that the internal chemical spread in nitrogen abundance, which 
is a signature of MPs, would not exceed 0.2 dex. Since Lindsay 113 is significantly older
 than other GCs with MPs, we suggest that the onset of MPs is likely determined by the 
 combination of cluster age and mass.
\end{abstract}


\section{Introduction} \label{S1}
The detection of chemical spread in almost all Galactic globular clusters 
(GCs, older than $\sim$10 Gyr) has made the notion of star clusters as 
simple stellar populations (SSPs) a view of the past. The star-to-star 
variations in the abundances of some light elements along different evolutionary 
phase \citep[e.g.,][]{Carr09a,Mari09a,Panc17a,Wang17a} in GCs strongly 
indicates that they are multiple stellar populations (MPs) rather than SSPs.  
{In photometry,} MPs are responsible for distinctive features along with the entire 
color-magnitude diagram (CMD), including multiple main sequences \citep[MSs,][]{Piot07a}, 
subgiant branches \citep[SGBs,][]{Vill07a}, red-giant branches \citep[RGBs,][]{Mari08a,Piot15a}, 
and also their combinations. \cite{Milo17a} has developed a 
pseudo-two-color diagram with a suitable combination of multiple passbands 
observed through the {\sl Hubble Space Telescope} ({\sl HST}). This diagram, 
which is called the ``chromosome map'' of GCs, is proved an
effective tool for detecting different stellar populations in GCs.

Numerous studies show that MPs are also present in extragalactic clusters. 
{\cite{Leta06a} have studied {detailed} chemical abundances among stars in three 
GCs of the Fornax dwarf spheroidal (dSph) galaxy. They find that the chemical 
pattern of Fornax dSph GCs is similar to those of Galactic GCs. This conclusion 
was underpinned by \cite{Lars14a}, in which they report the presence of 
nitrogen (N) variations among RGB stars for the same clusters, using the 
UV-optical-Infrared photometry based on the {\sl HST}.} MPs with different chemical 
composition were also detected in clusters in the Small and the Large 
Magelanic Cloud (SMC and LMC) by using both photometry \citep{Dale16a,Nied17a,
Mart18a,Lagi19a,Li19a} and spectroscopy \citep{Mucc09a,Holl19a}. 

To explain the origin of MPs, a significant fraction of models draw on self-pollution 
of the intra-cluster gas, which would be occurring during a cluster’s early evolution. 
Different scenarios invoke different stellar sources for chemical enrichment 
\citep[e.g.,][]{Decr07a,Derc08a,deMi09a}. However, detail explorations show that 
these scenarios would inevitably encounter problems when compare with 
observations \citep{Bast15a}. Recently, some scenarios also suggest a 
significant contribution from external accretion \citep{Li16a,Hong17a,Bekk19a}. 
All these scenarios would suggest that a significant fraction of gas content or 
young stellar objects can occur in advanced, massive clusters, which still 
remains controversial \citep[][however, \cite{For17a}]{Bast14a}.
 
One possible explanation is that stellar chemical anomalies are produced 
inside low mass stars and were present when they were sufficiently evolved 
\citep[e.g.][]{Bril89a,Cava96a,Cava98a}. The MPs in GCs might be 
caused by stellar evolutionary processes such as the first dredge-up and the 
Sweigart-Mengel mixing \citep{Swei79a}. The findings of chemical anomalies 
among less-evolved MS stars \citep[e.g.][]{Cann98a} have challenged 
these scenarios, because they are not hot enough for producing the observed 
chemical enrichment pattern. Alternatively, \cite{Jiang14a} proposed that stars 
with anomalous abundances in GCs might be the products of binary interactions. 
Recently \cite{Bast18a} suggest that MPs may be built 
with the help of some non-standard stellar evolutionary effects. They claim that  
a strong magnetic field associated with low mass stars may play a significant 
role in it, as only low mass stars that are magnetically braked exhibit 
obvious patterns of chemical variations. This scenario, although speculative, 
was supported by the evidence that clusters younger than $\sim$2 Gyr both 
in the Magellanic Clouds and in the Milky Way exhibit extended MS turnoffs 
\citep[e.g.,][]{Milo09a}, and clusters younger than $\sim$700 Myr 
show split or broadened MSs \citep[e.g.,][]{Li17a,Cord18a}. 

Clusters younger than $\sim$2 Gyr do not show evidence for light-element 
variations \citep[e.g.,][]{Mucc14a,Mart17a,Zhang18a}, hence they do not host MPs
with a different chemical composition. Indeed the eMSTOs and the multiple MSs of 
these young clusters are mainly due to the stellar rotation \citep[e.g.,][]{Yang13a,Dant17a,Geor19a}, 
as demonstrated by direct spectroscopic measurements of stellar rotation in MS and 
eMSTO stars \citep{Dupr17a,Mari18a,Mari18b,Sun19a}.

The ``self-enrichment'' or ``external accretion'' scenarios suggest 
that the total cluster mass would be the dominant property which defines the 
presence of MPs. On the other hand, ``stellar evolutionary'' scenarios 
would imply that MPs can only be observed in sufficiently evolved stellar systems. Therefore 
age is the key factor determining the occurrence of MPs. {For old GCs in the 
Milky Way, observations strongly indicate that the significance of the MPs depends on the 
cluster's total mass \citep[e.g.,][]{Carr10a,Milo17a,Milo18a}. In these studies, MPs' 
phenomenon systematically increases in both incidence and complexity with increasing 
cluster mass.} 

Only one old GC, Ruprecht 106 (Rup 106), was identified as a convincing sample of 
SSP GC \citep{Vill13a,Dott18a}. {However, MPs are also detected in GCs less 
massive than Rup 106 \citep[e.g., NGC 6535,][]{Brag17a}.} A similar correlation 
was also present in clusters with ages between 2 to 10 Gyr as well \citep{Chan19a}. 
According to \cite{Chan19a}, there seems to be a minimum threshold mass of 
about 30,000 $M_{\odot}$ for the appearance of MPs. However, the fact that their 
younger counterparts with comparable total masses do not exhibit MPs, has led to 
the argument that age is the major factor determining the presence of MPs.

Unfortunately, stellar populations in clusters of 2--6 Gyr-old 
are yet {less} studied. So far only {six} clusters in this age range were explored 
in terms of their stellar populations \citep{Mart18a,Chan19a,Holl19a,Li19a,Mart19a}. 
Among these clusters, NGC 1978, NGC 2121 and Hodge 6 were determined 
to have internal spread in their carbon (C) and N abundances, while Lindsay 113 was found 
to have a homogeneous distribution of helium abundance for its horizontal branch (HB) stars 
\citep[$\delta{Y}\leq0.01$ dex,][]{Chan19a}. 

In this work, we dissect the 4 Gyr-old SMC cluster Lindsay 113, 
which is a low mass cluster located well below the minimum threshold 
mass for clusters with MPs. We aim to examine if its RGB exhibit any 
intrinsic broadening which can be explained by internal chemical 
variations among its RGB stars. {Our work confirms that the observed 
width of the RGB of Lindsay 113 can be adequately explained by a SSP model, 
which disagrees with a recent study lead by \cite{Mart19a}, however.} 

In Section 2 we will introduce our data reduction. We present our analysis approach 
and main results in Section 3. {In Section 4 we will present the scientific implication 
and conclusions. We will also discuss the possible reason which leads to the 
disagreement between our result and that of \cite{Mart19a}.}


\section{Data Reduction} \label{S2}

The observations were obtained with both the {\sl HST}'s Ultraviolet and
Visual Channel of the Wide Field Camera 3 (UVIS/WFC3) and the Advanced 
Camera for Surveys/Wide Field Channel (ACS/WFC). The UVIS/WFC3 images 
were observed through the F343N and F438W passbands (program ID: 
GO-15062, PI: N. Bastian). The ACS/WFC images (program ID: GO-9891, 
PI: G. F. Gilmore) are collected through the F555W (480s) and F814W (280s) 
passbands.  The UVIS/WFC3 dataset includes three frames taken through 
the F343N passband, with exposure times of 540 s and 1065 s (twice), and 
three frames taken through the F438W passband, with exposure times of 128 s and 
545 s (twice). 

We used the specific photometric package designed for {\sl HST} observations, 
the {\sc Dolphot2.0} \citep{Dolp11a,Dolp11b,Dolp13a}, to perform the 
point-spread-function (PSF) photometry to the `{\tt \_flt}' and `{\tt
  \_c0f}' frames. We used the corresponding WFC3 and ACS modules to deal 
with images taken from different observational channels. {\sc Dolphot2.0} 
has integrated functions of charge-transfer efficiency corrections and 
photometric calibration routines such as aperture and zeropoint corrections 
when applying measurement to {\sl HST} observations, which make it 
a powerful tool for {\sl HST} photometry. 

After photometry, we have filtered the raw stellar catalogue by applying 
the following criteria: (1) identified by {\sc Dolphot2.0} as a `bright star'. 
(2) Not centrally saturated in any of each passband. (3) Sharpness is 
greater than $-$0.3 but less than 0.3. (4) Crowding parameter is less 
than 0.5 mag. Finally we reach two stellar catalogues derived from 
the UVIS/WFC3 and ACS/WFC observations, with total number of 
``good stars'' of 5,813 and 9,960, respectively. We then combine 
these two catalogues by cross-matching their spatial distributions, 
leading to a final stellar catalogue containing 4,380 stars.

\section{Main Results}
\subsection{Adopted Models}
\begin{figure*}[!ht]
\includegraphics[width=2\columnwidth]{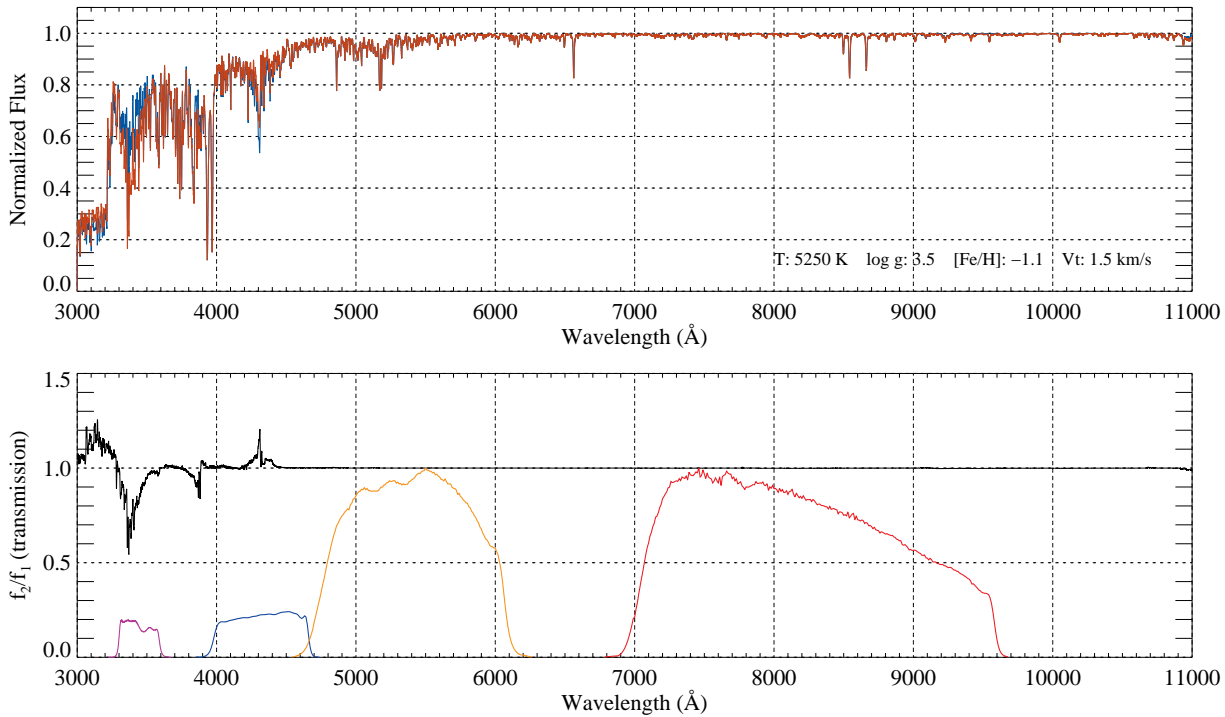}
\caption{Top: Model spectra of different CNO composition. 
The blue spectrum represents the star of ‘normal’ abundances, while the red one represents 
the star of enhanced nitrogen and depleted carbon and oxygen ($\Delta$[N/Fe] = +0.8~dex, 
$\Delta$[C/Fe] = $\Delta$[O/Fe] = -0.4~dex). The stellar parameters are indicated at the 
right-bottom corner.
Bottom: Flux ratio and filter transmission curves used here (from left to right: F343N/WFC3, 
F438W/WFC3, F555W/WFC, F814W/WFC).}
\label{F1}
\end{figure*}

The color {indices} build with appropriate combination of ultraviolet, optical and 
near infrared magnitudes, is an effective tool to search for MPs in GCs \citep{Lars14a}. 
In particular, the F343N passband is sensitive to the N abundance as 
it includes the molecular absorption band of NH ($\sim3370${\AA}). 
The F438W contains the CH absorption band ($\sim4300${\AA}), thus it 
is sensitive to C abundance. A typical second population star, which is 
usually N-enriched and C-depleted, will be fainter in F343N but 
brighter in F438W than a normal star with the same stellar atmospheric 
parameters ($\log{T_{\rm eff}}$, $\log{g}$, [Fe/H]). In addition, the 
temperature difference caused by different helium abundance between 
normal and enriched population stars could be revealed by the wide  
color baseline of F438W$-$F814W colors. 
We illustrate this principle in Figure \ref{F1} with the model spectra of 
one normal and one enriched star located at the base of the RGB with the stellar parameters 
of $T_{\rm eff} = 5250~K$, $\log{g} = 3.5$, [Fe/H] = -1.1~dex and $V_{\rm t} = 1.5~km~s^{-1}$. 
Keeping the total CNO abundance constant (i.e., $\Delta[(C+N+O)/Fe]=0$ dex), the carbon, nitrogen 
and oxygen abundances of enriched stars are set to be [C/Fe] = [O/Fe] = $-$0.4~dex and 
[N/Fe] = +0.8~dex. Our model spectra are synthesized with the iSpec \citep{Blan14a,Blan19a} 
where the radial transfer code SPECTRUM \citep{Gray94a} 
and the MARCS model atmosphere \citep{Gust08a} are adopted.

\begin{figure*}[!ht]
\includegraphics[width=2\columnwidth]{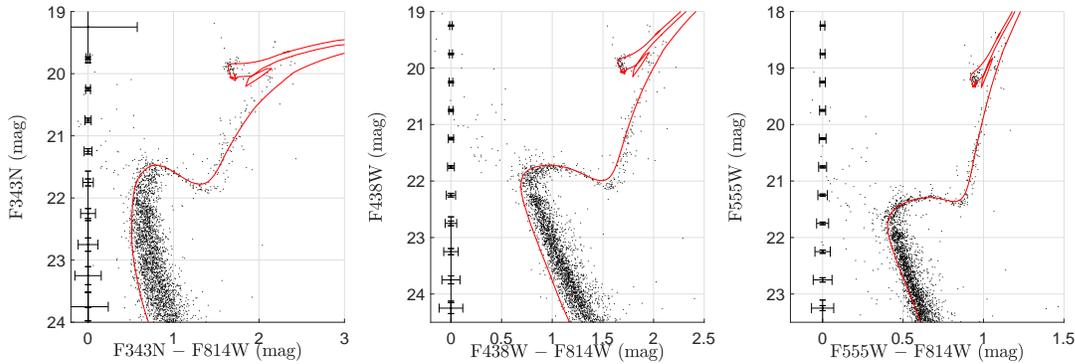}
\caption{{Lindsay 113} CMDs. (left) F343N$-$F814W vs F343N; (middle)
  F438W$-$F814W vs F438W; (right) F555W$-$F814W vs F555W. Red
  lines are best-fitting isochrones. In each panel, average photometric 
  uncertainties (corresponding to 90\% confidence interval) are on the left side.}
\label{F2}
\end{figure*}

In Figure \ref{F2} {we show three CMDs in different pass-bands of the Lindsay 113 cluster.}
We then fit these CMDs through the MESA Isochrone and Stellar Tracks \citep[MIST;][]{Paxt11a,Paxt13a,Paxt15a,Choi16a,Dott16a}. As previously noticed by \cite{Bark18a,Li19a}, 
we are not able to provide simultaneously a good fit to all CMDs. The poor fit is likely 
due to the uncertainties on the adopted extinction law in the SMC\footnote{MIST model 
sets the Milky Way extinction curve as the sole option when interpolate isochrones}.  
We find the isochrone can fit the (F555W, F555W$-$F814W) CMD very well. 
However, the fitting deteriorates when the first passband becomes bluer. 
Because of this, in this work, we only focus on the width of the RGB rather 
than its position. Moreover, we adopt for Lindsay 113 the age, distance and metallicity 
inferred from the (F555W, F555W$-$F814W) CMD, where the effect of reddening is less 
evident, which are, $\log{(t\;{\rm yr}^{-1})} =9.64\pm0.02$ 
(4.37$\pm$0.20 Gyr), [Fe/H] = $-1.15\pm0.10$ dex, $(m-M)_0 =18.78\pm0.05$ 
mag and $E(B-V)$=0.05$\pm$0.01 mag. The associated uncertainties 
were defined by the size of the adopted grids. We determined the best-fitting 
age by inspecting the relative position of the isochrone to the SGB. The 
metallicity was determined by the slope of the RGB. The best-fitting distance 
modulus, as well as the extinction value, were defined by the position of the 
red clump (RC). For another two CMDs, the fitting of the isochrones to the 
positions of the RGB and the RC is our priority, because we are more 
interested in these parts rather than its MS. Our results are similar to those 
found in \cite{Piat18a,Chan19a}.

\subsection{Synthetic SSPs}
Based on the best-fitting MIST isochrone, we simulated multi-bands photometry 
to compare with our observations. Possible effects other than the star-to-star 
chemical variations which may broaden the RGB include (i) photometric 
uncertainties and artifacts (e.g., cosmic rays, hot pixels), (ii) different 
distances of individual stars, (iii) differential reddening, and (iv) overlapped field 
stars with different ages and metallicities. The potential broadening of RGB caused 
by different stellar distances is negligible because of the large distance 
of the cluster. {In principle}, by using the artificial star technique, we can evaluate the 
effects of photometric uncertainties and artifacts. To do this, we have generated more than 
one million artificial stars (ASs) located on the best-fitting isochrone. We defined 
10,000 sub-sample of ASs contaning 100 stars each and added them to the images 
by using the appropriate PSF model derived from real stars.  These stars have been 
reduced by using the same method adopted for real stars. Our final artificial catalog 
contains 751,310 stars. We verified that most of the non-detected ASs are either 
faint stars or saturated objects. From these 751,310 stars, we randomly selected 
roughly the same number of artificial stars to the observation based on the observed 
luminosity function. 

{However, for the real observation, additional uncertainties may still occur due to multiple 
reasons. First, the PSF-fitting would never be perfect for real stars, {as it would for the 
simulated ASs.}  Also, even there is no differential reddening effect, the photometric zero 
points would still exhibit spatial dependent variations due to the inaccuracies in the 
determination of the background and/or in the charge transfer efficiency correction. 
In principle, these effects are small but unpredictable. \cite{Milo12a} have compared 
the difference between the color spread of observed MS stars in GCs (with differential 
reddening corrected) with that of ASs. Tthey find that the observed width of MS is about $\sim$0.003--0.005 mag wider than the artificial MS, indicating an additional magnitude 
spread of 0.002--0.004 mag on average. In this work, we have also added additional noise 
of $\delta$=0.003 mag in each passband. We argue that any conclusion on the presence of 
multiple populations in Lindsay 113 (or any other cluster) should account for these 
properties of ASs.} 

{Recently \cite{Mart19a} conclude that the effect of differential reddening in 
this cluster is negligible. However, we do find that the observed patterns in the observed 
CMD are more dispersed than those of the ASs without differential reddening (even taken 
the additional noise into account as described above). We conclude that Lindsay 113 is 
indeed affected by differential reddening. This is the {\sl major} difference between 
\cite{Mart19a} and this work. We will provide more details in Section \ref{S4p1} for this issue.}

\begin{figure*}[!ht]
\includegraphics[width=2\columnwidth]{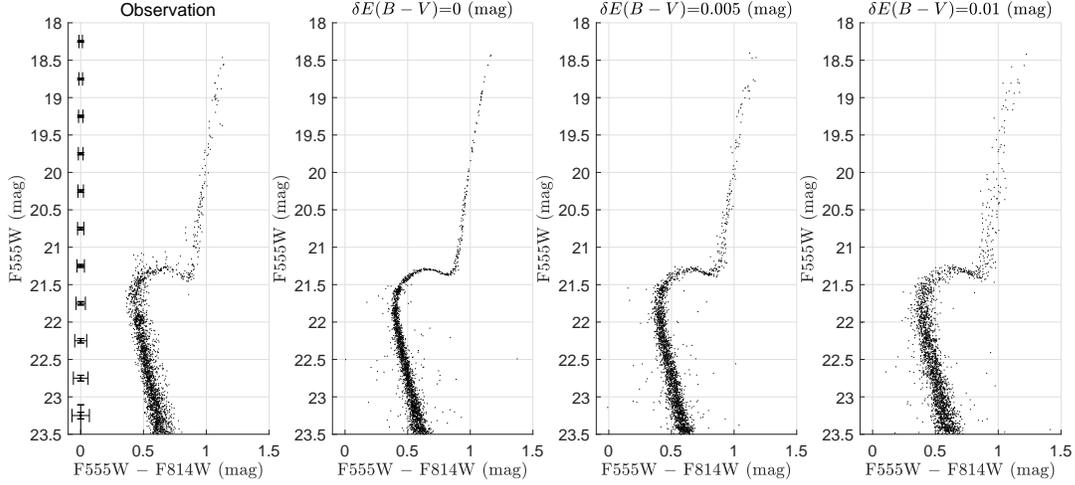}
\caption{The observed CMD of Lindsay 113 with error bars (left) and simulated CMDs 
characterized by different degrees of differential reddening (as indicated by their titles). 
The added differential reddening was derived from a Gaussian distribution.}
\label{F3}
\end{figure*}

The effect of differential reddening was explored by comparing the simulated CMDs with 
the observational CMD in F555W and F814W, because the effect of possible N spread 
will not {significantly} affect the photometry in these passbands. In addition, \cite{Chan19a} 
have concluded that the helium variation among Lindsay 113 stars is negligible. We find 
that the observed CMD is indeed more broadened in all parts (MS, SGB, RGB) than the 
simulated CMD without differential reddening. {We then compare the width of the 
observed RGB with that of the simulation with differential reddening (see Section \ref{S3p3} 
for the identification of the RGB). We conclude that the degree of differential reddening is 
most likely $\delta{E(B-V)}=0.005\pm0.002$ mag.} Figure \ref{F3} shows the observed CMD 
(left) as well as three simulated CMDs with differential reddening degrees of 
$\delta{E(B-V)}=0.0,0.005$, and 0.01 mag, 
respectively. From Figure 3, one can see that Lindsay 113 does not have an eMSTO, as 
expected \citep{Geor19a}.

\subsection{Statistical Analyses}\label{S3p3}
\begin{figure*}[!ht]
\includegraphics[width=2\columnwidth]{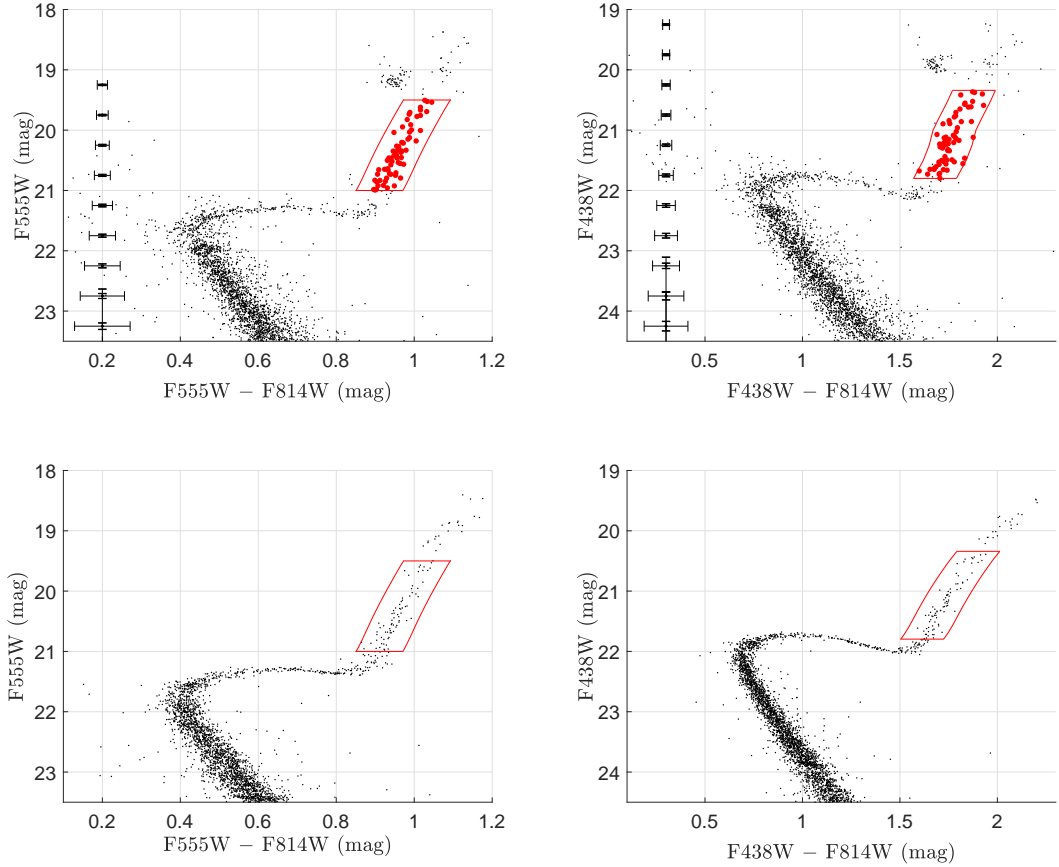}
\caption{Identification of RGB stars from the observations (with error bars 
on the left side). Stars located 
in selection boxes of both CMDs involving F438W, F555W and F814W 
are member RGB stars. The selection boxes were determined based 
on the artificial CMDs (bottom panels).}
\label{F4}
\end{figure*}

Figure \ref{F4} illustrates how do we select RGB stars. We used the simulated 
CMDs in F438W and F555W in respect with F814W 
to determine the regions occupied by the majority of RGB stars. We determined 
the RGB selection boxes by examining their relative deviation to the ridgeline 
(for the simulated RGB, the ridgeline is the best-fitting isochrone). Only stars 
located in both selection boxes were considered RGB stars. This method has been  
used for minimizing the effect of field-star contamination \citep{Mart17a}.
We have identified 67 candidates as member RGB stars. 

\begin{figure*}[!ht]
\includegraphics[width=2\columnwidth]{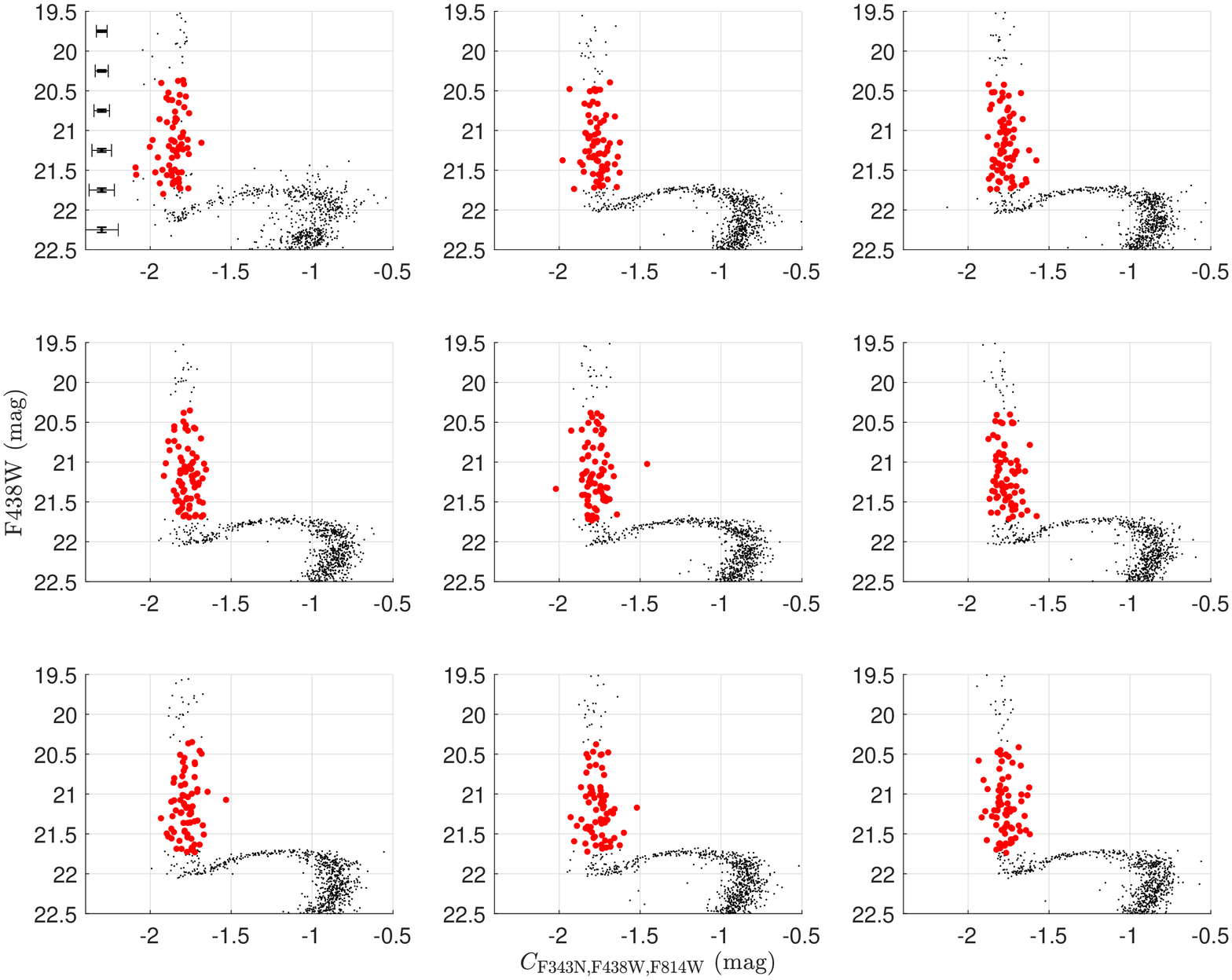}
\caption{The $C_{\rm F343N,F438W,F814W}$ vs. F438W diagram 
for the observation with error bars (first row, left panel), the red dots are selected 
RGB stars. Other panels exhibit eight examples of simulated SSPs}
\label{F5}
\end{figure*}

We constructed a color index, 
$C_{\rm F343N,F438W,F814W}$ = (F343N$-$F438W)$-$(F438W$-$F814W), 
to quantify the broadening of the RGB caused by chemical 
variations, which is similar to the index used in \cite{Mone13a}. This 
pseudo-color index is sensitive to unveil internal spread of C and N 
abundances \citep{Mone13a,Mart17a}. We then compared the observed 
RGB in the $C_{\rm F343N,F438W,F814W}$ versus F438W diagram with 
that for artificial RGB. In Figure \ref{F5} we compare the observed CMD with eight 
simulated CMDs of SSPs derived from randomly-extracted ASs. We did not see 
any visible difference between the observation and simulations (SSPs). The 
analysis of a large sample of over 100 simulated CMDs confirm that the observed 
RGB width is consistent with the width expected from observational errors alone.

\begin{figure*}[!ht]
\includegraphics[width=2\columnwidth]{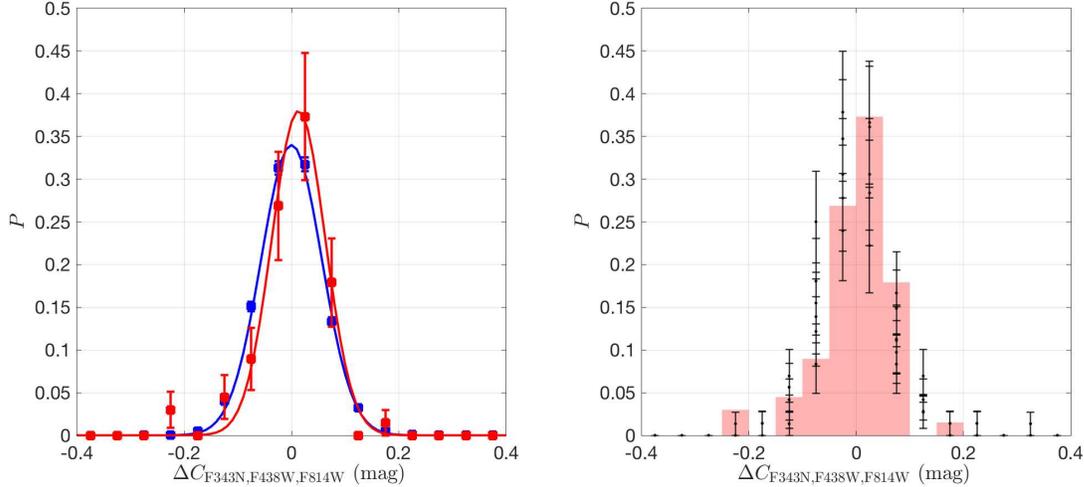}
\caption{Left: the observed distributions of $\Delta{C_{\rm F343N,F438W,F814W}}$ 
(red circles) versus that for the simulations (blue circles). The red and blue curves 
are their best fitting Gaussian curves. Right: the observed histogram for stars 
with different $\Delta{C_{\rm F343N,F438W,F814W}}$. The black dots with error bars 
are the $\Delta{C_{\rm F343N,F438W,F814W}}$ distributions for eight individual 
simulated samples (corresponds to Figure \ref{F5}).}
\label{F6}
\end{figure*}

In Figure \ref{F6} we compared the $\Delta{C_{\rm F343N,F438W,F814W}}$ 
distributions (histograms) of observed RGB with those of simulated CMDs. Here $\Delta{C_{\rm F343N,F438W,F814W}}$ is defined as the relative deviation of $C_{\rm F343N,F438W,F814W}$ 
for individual stars to their average $C_{\rm F343N,F438W,F814W}$ of stars with 
similar magnitude. We find that the observed distribution of $\Delta{C_{\rm F343N,F438W,F814W}}$ 
for RGB stars is well consistent with that of the simulations, i.e., SSPs. We have 
fitted the distributions of the pseudo-color index for both the observation and the average 
distributions of the simulated CMD, using a Gaussian profile. For the simulations, their average 
distribution of the $\Delta{C_{\rm F343N,F438W,F814W}}$ can be described as 
follow, 
\begin{equation}
P(\Delta{C})=0.34e^{-\left({\frac{\Delta{C}}{0.08}}\right)^2},
\end{equation}
while for the observation, it is
\begin{equation}
P(\Delta{C})=0.38e^{-\left({\frac{\Delta{C}-0.01}{0.07}}\right)^2},
\end{equation}
this internal spread of the pseudo-color index ({corresponding to a 68\% confidence}) 
for the observed RGB stars is $\sigma=0.07\pm0.02$ mag\footnote{The uncertainty 
was estimated by a bootstrap test.}, which is consistent with the simulated SSPs 
(${\sigma=0.08\pm0.02}$). {In this example we have used 5000 ASs. }
These results are shown in Figure \ref{F6}.

{For the analysis above, we compare the observed color distribution with the color
distribution of ASs, by adding to ASs a reasonable amount of differential reddening. 
As an alternative, we also studied the color distribution of the RGB corrected for 
differential reddening and compared it with the color distribution of ASs. In this case 
we did not add any differential reddening to ASs, but only a reasonable 
{error} associated to the differential reddening. 

To investigate the effect of differential reddening on the diagrams of
Lindsay 113 we applied the method by \cite{Milo12a} to the
F336W vs. F336W$-$F814W CMD, which is the CMD that provides the wider
color baseline. We selected a sample of $\sim$1000 well-measured 
bright MS stars and derived the fiducial line. We derived for each star its
residual from the fiducial, along the reddening line, and then corrected
the star's color and magnitude by the median residual of its 45
selected MS neighbors. As a result, we find that $\sim$97\% of stars in the 
field of view are affected by a reddening variation of $-0.016\leq\Delta{E(B-V)}\leq0.016$  
mag, while 68\% of our sample have a differential reddening of $-0.006\leq\Delta{E(B-V)}\leq0.006$. 
This further {supports} that our previous adoption of $\delta{E(B-V)}$=0.005$\pm$0.002 mag to 
the ASs is reasonable. 

In Figure \ref{F7}, we show the performance of our correction of differential 
reddening. We selected two subsamples with $\Delta{E(B-V)}>0.005$ mag and 
$\Delta{E(B-V)}<-0.005$ mag. We then plot their CMDs (F336W vs. F336W$-$ F814W) 
together (the left-top panel of Figure \ref{F7}). One can see that stars with $\Delta{E(B-V)}>0.005$ 
mag are indeed systematically redder than stars with $\Delta{E(B-V)}<-0.005$ mag. 
In the left-bottom panel of Figure \ref{F7}, we show the same CMD with differential 
reddening corrected. As a result, we find that the color difference between 
these two subsamples disappears, and all parts of the CMD {become} much narrower 
than those of the raw CMD. In the right panel, we plot the differential reddening map 
in the whole field of view. 

\begin{figure*}[!ht]
\includegraphics[width=2\columnwidth]{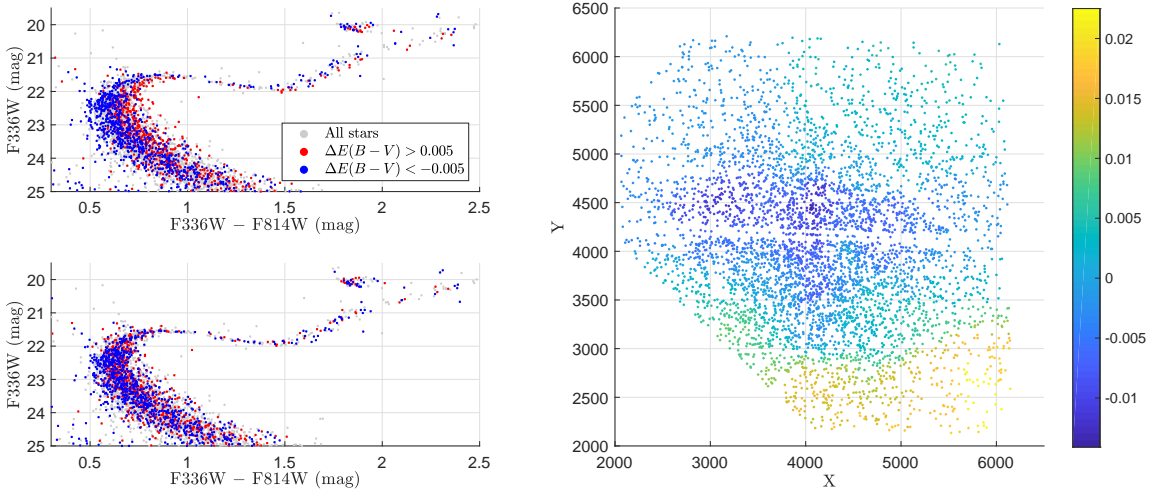}
\caption{Left-top: CMDs of stars with $\Delta{E(B-V)}>0.005$ mag and $\Delta{E(B-V)}<-0.005$ mag. 
Left-bottom: CMDs of stars with $\Delta{E(B-V)}>0.005$ mag and $\Delta{E(B-V)}<-0.005$ mag, with 
differential reddening effect corrected. Right: the differential reddening map ($\Delta{E(B-V)}$ (mag))  for all stars.}
\label{F7}
\end{figure*}

Because the method we used for correcting the differential reddening is still based on 
statistics, each value of differential reddening has associated an error of $\sigma/N$, 
where $\sigma$ is the dispersion of the residuals of the N neighbors used for 
the correction. The average associated errors in our case is 
$\delta{E(B-V)}\sim$0.0026 mag. Similar to previous analysis, in Figure \ref{F8} 
we show a comparison between the differential reddening corrected diagram for the 
observation and the ASs. {The AS sample was incorporated with noise from photometric 
uncertainty, artifacts and position-dependent zero point, as well as the variations of PSF-fitting 
and reddening residuals. But in this case, it is free of differential reddening.} 
We confirm that the observed RGB in the $C_{\rm F343N,F438W,F814W}$ 
vs. F438W diagram becomes narrower and similar to that of ASs without differential 
reddening.

\begin{figure*}[!ht]
\includegraphics[width=2\columnwidth]{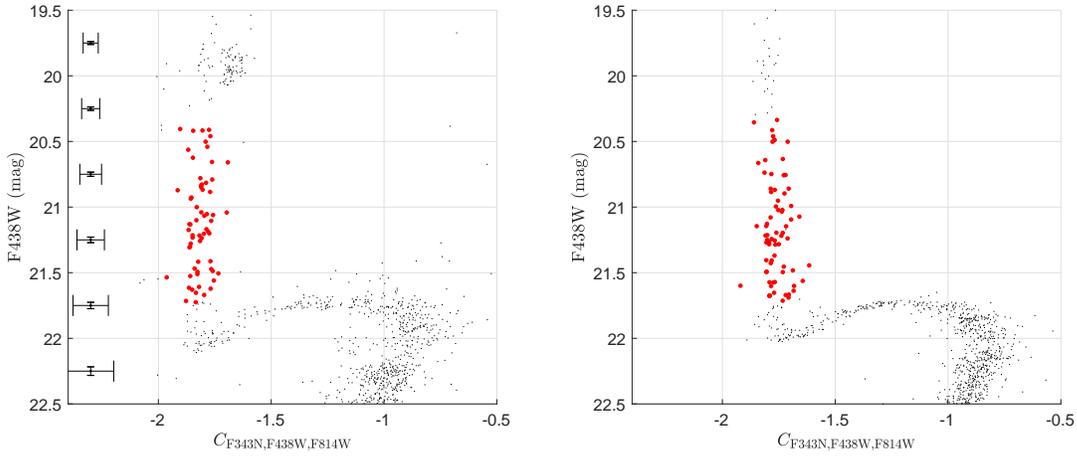}
\caption{Similar to Figure \ref{F5}, but the observed $C_{\rm F343N,F438W,F814W}$ vs. F438W diagram (left) was corrected for differential reddening. The same diagram for an example of ASs that is free of differential reddening, is presented in the right panel.}
\label{F8}
\end{figure*}

Similar to Figure \ref{F6}, in Figure \ref{F9} we plot the distributions of $\Delta{C_{\rm F343N,F438W,F814W}}$ for both the observed RGB stars with differential reddening 
 corrected, and the ASs without differential reddening effect. Again, we used a single 
Gaussian function to fit these distributions: for the simulated ASs, the best fitting Gaussian 
function is
\begin{equation}
P(\Delta{C})=0.46e^{-\left({\frac{\Delta{C}}{0.06}}\right)^2},
\end{equation}
and for the observation, it is 
\begin{equation}
P(\Delta{C})=0.48e^{-\left({\frac{\Delta{C}}{0.06}}\right)^2},
\end{equation}

For both the observation and the simulation, their RGB stars all exhibit an internal 
spread of the pseudo-color index of $\sigma=0.06\pm0.02$ mag. Again, this result 
supports that there is no significant chemical variation among the RGB stars of Lindsay 113. 

\begin{figure}[!ht]
\includegraphics[width=\columnwidth]{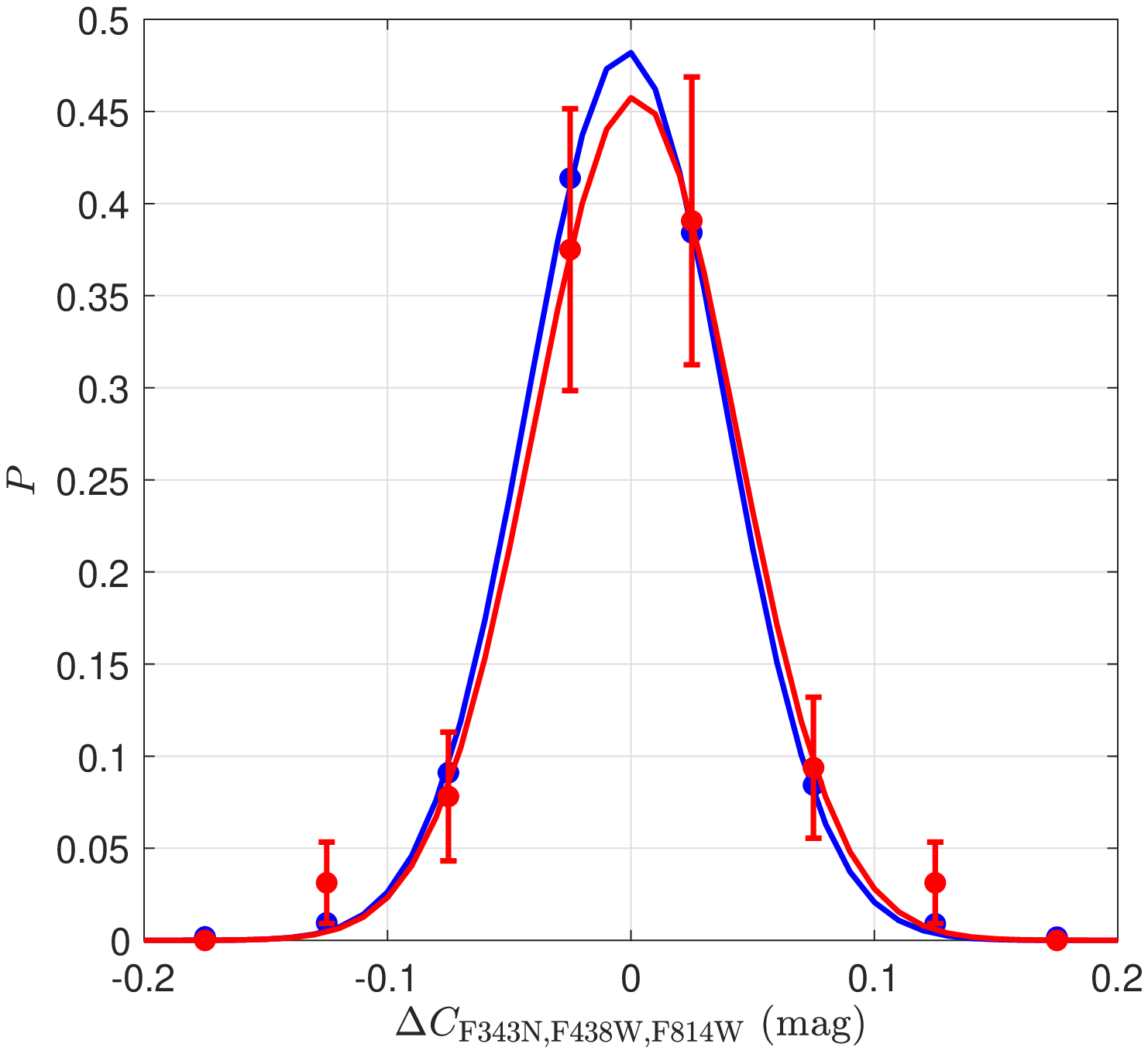}
\caption{The same as the left panel of Figure \ref{F6}, but the observed RGB 
stars have their differential reddening corrected (indicated by red circles), while the simulated ASs are differential reddening free (blue circles).}
\label{F9}
\end{figure}
}

\subsection{Constraint on the internal chemical variation}
\begin{figure*}[!ht]
\includegraphics[width=2\columnwidth]{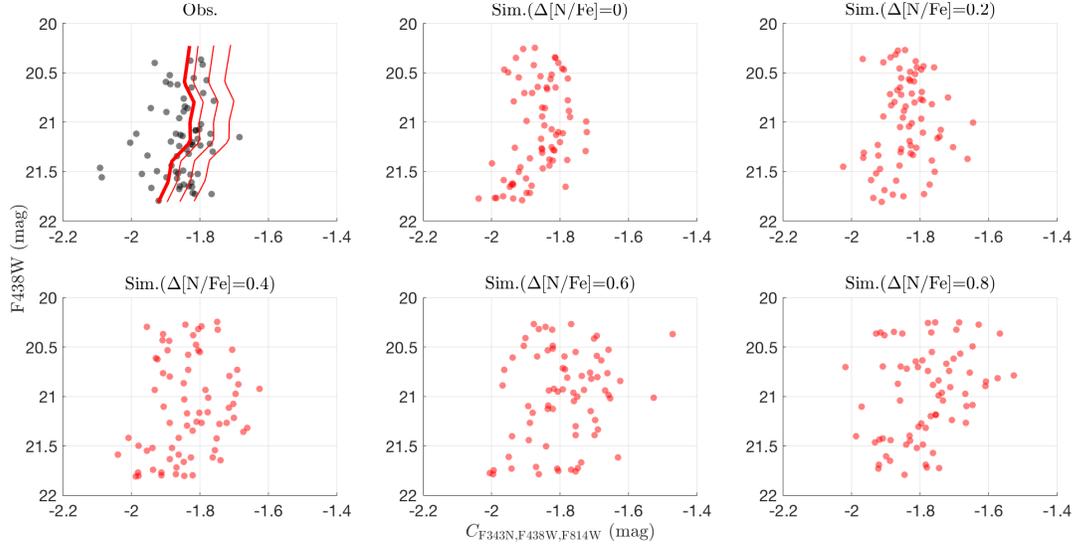}
\caption{The observed (top-left panel) and the simulated RGB stars with different CNO 
abundances in the $C_{\rm F343N,F438W,F814W}$ versus F438W diagram. The thick solid 
line is the best fitting ridgeline to the observation. From left to right, the thin solid lines 
are loci with $\Delta$[N/Fe]=0.2, 0.4 and 0.6, respectively. Only for models with 
$\Delta$[N/Fe]=0.0 and 0.2 (top-middle and top-right), the two-sample K-S test will report 
that the simulated RGB stars have their $C_{\rm F343N,F438W,F814W}$ distributions 
similar to the observation (over 80\% of simulations have reproduced the observation). 
For $\Delta$[N/Fe]=0.4, only $\sim$24\% of simulations can reproduce the observation. 
For $\Delta$[N/Fe]=0.6, this ratio rapidly decreases to $\sim$1\% (see Table \ref{T1})}
\label{F10}
\end{figure*}

Using the {MARCS} model atmospheres \citep{Gust08a}, based on the best 
fitting isochrone, we calculated the corresponding loci with different 
C, N and O abundances. We calculate the relative deviation 
between loci with $\Delta$[N/Fe]$=$0.2, 0.4, 0.6 and 0.8 dex to the standard isochrone. 
We then apply these deviations to the ridgeline of the observed RGB in the ${C_{\rm F343N,F438W,F814W}}$ vs. F438W diagram. Finally we obtained four ridgelines 
for populations with abundance variations as introduced above. Based on these 
ridgelines, we simulated one SSP and four MPs with different [N/Fe]. For each 
simulated population, the number of stars is equal to the observation (readers can 
direct to our Figure \ref{F10} for a quick view). 

To quantify the similarity between the observed and simulated distributions of the 
$C_{\rm F343N,F438W,F814W}$, we have applied a two-sample Kolmogorov–Smirnov 
test (K-S test) to the observation and each simulation. For each run, the K-S test will return 
two values, $H$ and $P$. The outcome of $H$ depends on the null hypothesis 
that the two underlying distributions (the observation and tested simulation) 
are independent. If $H$ returns 0, that means the null hypothesis is rejected, indicating 
that the observed and simulated $C_{\rm F343N,F438W,F814W}$ are drawn from the 
same distribution. Otherwise, it returns 1. The $P$ represents the probability that the 
two samples are drawn from the same underlying population. For an individual 
K-S test, typically when the $P$ value is higher than 0.05, the null hypothesis 
is rejected.  

To avoid the uncertainty caused by the small number statistics, for each simulation, 
we repeat the procedure 10,000 times. We then examine how many times 
the two-sample K-S test will return a $H$=1 and calculate the average $P$ value 
for these 10,000 runs. Our results are summarized in Table \ref{T1}. 

\begin{table}
 \centering
  \caption{The outcomes of the two-sample K-S test between the observed and 
  simulated $C_{\rm F343N,F438W,F814W}$ distributions.}\label{T1}
  \centering
  \begin{tabular}{@{}lcc@{}}
  \hline
        & $N(H=1)\footnote{We have totally ran 10,000 times.}$ & $\bar{P}$ \\
 \hline
 $\Delta$[N/Fe]=0.0 (SSP) & 9084 & 0.31 \\
 $\Delta$[N/Fe]=0.2 & 8092 & 0.23 \\
 $\Delta$[N/Fe]=0.4 & 2403 & 0.04 \\
 $\Delta$[N/Fe]=0.6 & 94 & 2.6$\times10^{-3}$ \\
 $\Delta$[N/Fe]=0.8 & 0 & 7.3$\times10^{-5}$ \\\hline
\end{tabular}
\end{table}

As shown in Table \ref{T1}, we find that for models with $\Delta$[N/Fe]=0.0 
or  $\Delta$[N/Fe]=0.2, most runs (over 8000) of the two-sample K-S test will report 
$H$=1. When increasing the internal nitrogen variation of models to $\Delta$[N/Fe]=0.4, 
the number of runs with $H$=1 sharply decrease to 2403, and their average $P$ 
value also decreases to 0.04. For models with $\Delta$[N/Fe]=0.6 and 0.8, only 
94 and zero runs will report $H$=1. Therefore we conclude that the internal chemical 
spread of the observed RGB stars is most likely $\Delta$[N/Fe]$\leq$0.2. 
In Figure \ref{F10}, we exhibit examples of the observed and simulated RGB in 
the $C_{\rm F343N,F438W,F814W}$ vs. F438W diagrams. Among these examples, only for 
models with $\Delta$[N/Fe]=0.0 and 0.2, the K-S test returns $H$=1. For other models 
with $\Delta$[N/Fe]=0.4, 0.6 and 0.8, the K-S test returns $H$=0. 

\section{Discussion and Summary}
\subsection{A possible solution to the disagreement to \cite{Mart19a}}\label{S4p1}
{Recently \cite{Mart19a} have studied four intermediate-age clusters 
include Lindsay 113. In contrast to this work, they conclude that Lindsay 113 
{harbors} MPs because of the observed broadening of the RGB in this cluster. 
Our work confirms the evidence of a color broadening in the RGB of
Lindsay 113; however, our analysis suggests that such broadening is mostly
due to differential reddening. 

One reason that may explain the disagreement between this work and  
\cite{Mart19a} is that they have used {a sample of RGB stars, which are less appropriate 
to correct the differential reddening. In fact,} the total number of stars from the base 
to the top of the RGB is only $\sim$150, that means for each particular star, they cannot find a sufficiently large number of nearby stars to study the average reddening in a small 
spatial cell. {We suggest that the spatial resolution of their differential reddening map is 
poor, that might explain} they claim that the effect of differential reddening effect 
is negligible. Indeed, as shown in Figure \ref{F3} and Figure \ref{F7}, 
this effect is significant. 

The differential reddening effect is important in particular for observations involving 
UV passbands. \cite{Mart19a} have studied the distribution of RGB stars in the 
color index of $C_{\rm F343N,F438W,F336W}$\footnote{(F343N$-$F438W)$-$(F438W$-$F336W)}, 
for which two UV passbands ({F336W and F343N}) are involved. As a result, they find that the 
spread in this color index between the observed RGB stars and ASs is a little different, which 
are $\sigma=0.041\pm0.003$ mag and $\sigma=0.024$ mag.\footnote{here $\sigma$ 
indicates the standard deviation of $C_{\rm F343N,F438W,F336W}$ for RGB stars.} 
As shown in this work, the average differential reddening degree is most likely 
$\delta{E(B-V})$=0.005 mag, which will lead to an additional uncertainties of 
$\delta{F336W}\approx\delta{F343N}$=1.6$\times$3.1$\times$$\delta{E(B-V)}$=0.025 mag, 
and $\delta{F438W}$=1.3$\times$3.1$\times$$\delta{E(B-V)}$=0.02 mag. These 
uncertainties will finally combine into uncertainty of $\delta{C_{\rm F336W,F438W,F343N}}$=0.045 mag. The differential reddening is indeed important.

{Using the same} method and database utilized in \cite{Mart19a}, we calculated 
the $\Delta{C_{\rm F336W,F438W,F343N}}$ distribution for a differential reddening free AS 
sample. The standard deviation of these artificial RGB stars is 
$\delta{(C_{\rm F336W,F438W,F343N})}$=0.027 mag, which is consistent with \cite{Mart19a}. 
We then assume a degree of $\delta{E(B-V)}$=0.005 mag for all ASs, afterthat the artificial 
RGB becomes wider with $\delta{(C_{\rm F336W,F438W,F343N})}$=0.052 mag, which fully 
explains the observed width of RGB stars in $C_{\rm F336W,F438W,F343N}$ 
($0.041\pm0.003$ mag). Our result is presented in Figure \ref{F11}.

\begin{figure}[!ht]
\includegraphics[width=\columnwidth]{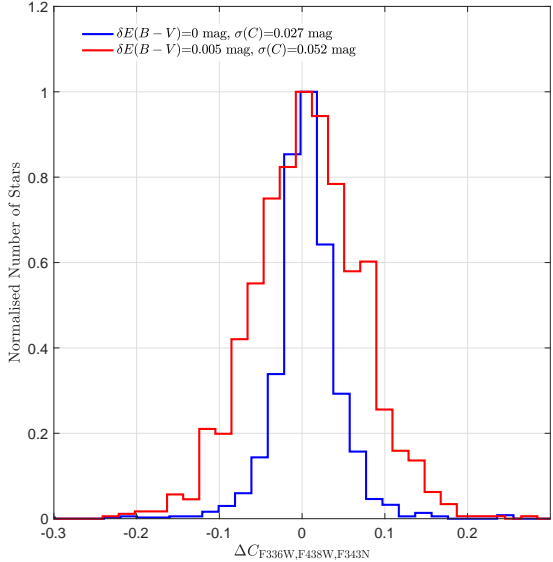}
\caption{The $\Delta{C_{F336W,F438W,F343N}}$ distribution for artificial RGB stars 
with/without differential reddening (red/blue histograms).}
\label{F11}
\end{figure}

Therefore, {we suggest that not taking differential reddening into account might hamper} the conclusion about Lindsay 113 made by \cite{Mart19a}. No evidence of significant chemical variation can be detected through this photometric database.
}

\subsection{Scientific Implications}
\begin{figure*}[!ht]
\includegraphics[width=2\columnwidth]{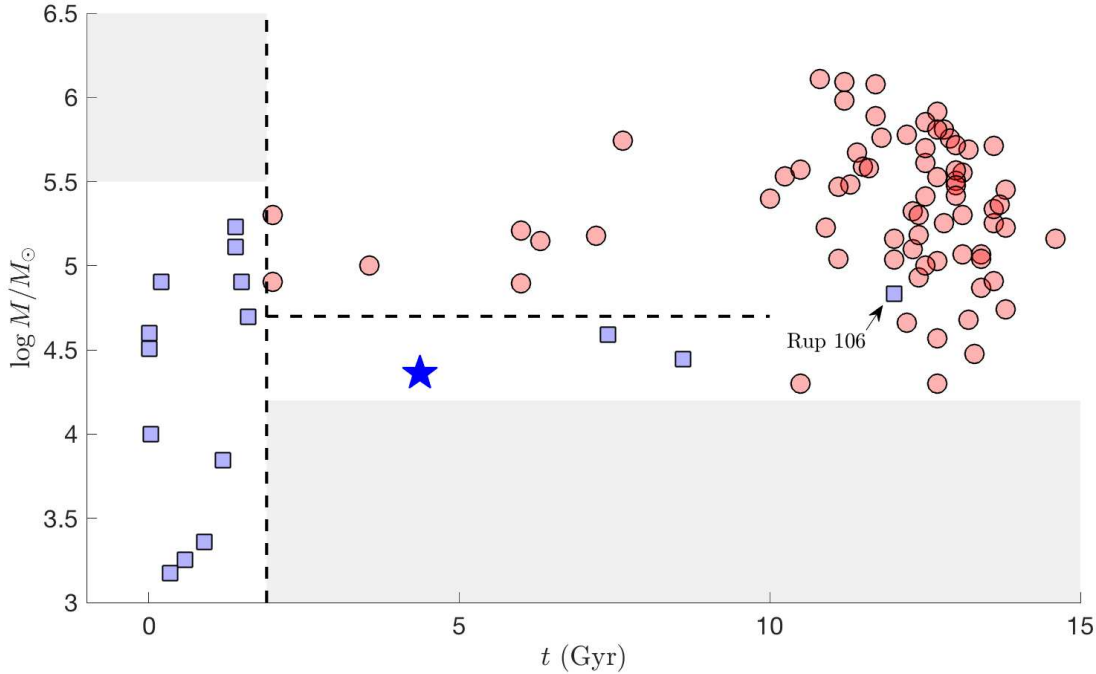}
\caption{The age-mass plane for clusters with and without MPs (red circles and 
blue squares), and Lindsay 113 (the blue pentagram). The vertical and horizontal dashed 
lines indicate the possible age and mass boundary which may define the presence 
of MPs. Grey areas were the suggested searching parameter space for clusters 
in future studies. {Data were obtained from \cite{Mcla05a,Baum13a,Krau16a}.}}
\label{F12}
\end{figure*}

{In analogy to \cite{Mart18a}}, we summarize clusters with and without MPs into the 
age--mass diagram in Figure \ref{F12}. There is a clear age boundary located between 
1.8 Gyr and 2.0 Gyr, as indicated by the vertical dashed line. 
All clusters younger than this age range do not exhibit MPs, while most 
of their older counterparts do. For clusters between 2 to 10 Gyr, it seems 
that low mass clusters (less massive than $\sim$50,000 $M_{\odot}$, 
horizontal dashed line) do not have MPs. However, it remains quite uncertain 
because the number of studied stars in these clusters are small 
\citep[Terzan 7,Palomar 12,][]{Taut04a,Sbor07a}, for which only three and 
five stars were studied through high-resolution spectra. For Milky Way GCs, 
only one cluster, Ruprecht 106 (the arrow in Figure \ref{F12} indicates its 
position in the mass-age plane), lacks MPs, as reported by \cite{Vill13a} based on 
studies of nine member stars. This work is verified recently by \cite{Dott18a} as well. 
{However, because {MPs have been detected in lower mass GCs} \citep[e.g.,][]{Brag17a}, 
it seems Rup 106 does not set a strong constraint on the mass limitation for 
the presence of MPs.} Future studies of a larger sample of stars are essential 
for determining if there is a mass boundary for evolved clusters to harbor MPs.

In this paper, the photometric appearance of the intermediate-age ($\sim$4.4 Gyr), 
low mass ($\sim$23,000 $M_{\odot}$) cluster, Lindsay 113, was analyzed. 
We have studied its RGB morphology in the diagram of F438W versus 
$C_{\rm F343N,F438W,F814W}$, an effective tool which can unveil significant 
variations in C and N among RGB stars. We have carefully corrected the effect 
of photometric uncertainties, artifacts, and potential differential reddening. 
We do not find apparent broadening of the RGB when comparing with that of 
simulated SSPs. Our observation is consistent with a model characterized by internal 
C and N abundances {not exceeding} $\Delta$[C/Fe]=$-$0.1 dex and $\Delta$[N/Fe]=0.2 
dex. The {noise contributed by photometric uncertainties, artifacts and differential 
reddening, as well as} small number stochastic can adequately explain 
the observed width of the RGB.

The lack of MPs in Lindsay 113 provides solid evidence which 
indicates that clusters masses may still play a role on the presence of MPs, 
at least it defines the strength of the chemical spread among their member 
stars. Our result is in apparent contrast to the massive counterpart of Lindsay 113, 
NGC 2121: using the same method, \cite{Li19a} conclude that NGC 2121 would 
harbor an internal N spread of $\sim$0.5 dex. 

Our result is further supported by the fact that Lindsay 113 lacks an apparent helium 
variation among its RC stars \citep{Chan19a}. Detail studies based on high-resolution 
abundances for individual stars would be essential to define whether this cluster is 
SSP.  Studying this cluster in passbands of shorter wavelength 
(i.e., F275W at UVIS/WFC3) would provide additional evidence regards whether 
this cluster is a SSP, because MPs would exhibit variation in Oxygen (O) that can 
be observed through the OH molecular absorption at $\sim$2800$\AA$\citep[e.g.][]{Milo17a}. 

{It would be interesting to study stellar populations in younger 
($\leq$2 Gyr) and massive ($\geq$few $\times 10^5 M_{\odot}$) clusters, as 
well as older ($\geq$2 Gyr) but low mass ($\geq30,000 M_{\odot}$) clusters,  
as indicated by the grey areas in the age-mass plane (Figure \ref{F12}).} However, 
the Milky Way and its nearby satellite galaxies lack super young massive clusters, 
therefore to shed light on the MPs problem, we need next-generation telescopes to 
observe distant star forming galaxies. As an example, \cite{Smit06a} have 
detected five supermassive cluster candidates in the M82 galaxy, one of 
which has an estimated mass of over $10^6 M_{\odot}$ and an age of 
only $\sim$6 Myr. To resolve individual RGB stars in clusters at this distance, an at 
least 8-meter UV-optical space telescope is required. However, current ground-based 
telescopes with the equivalent aperture (such as the Very Large Telescope) would 
suffer problems in adaptive optics when correcting the atmospherical turbulence at this 
short wavelength. Therefore, a more feasible direction is to explore stellar populations 
among those older, low mass clusters. The Magellanic Clouds already have a handful 
of these clusters \citep[e.g., NGC 2193,][]{Baum13a}. However, one problem is that 
these low mass clusters may not have a well-populated RGB because of 
the small number of stars, making the photometric method problematic. 
Overall, a combination of both the photometric and spectroscopic methods is essential.
{Another feasible project is to explore whether the young and intermediate-age clusters 
would have O variations among their MK dwarfs. These O enhanced, late-type stars will 
exhibit strong absorption bands from the ${\rm H_2O}$ molecules, which can be detected 
in near-Infrared passbands. However, given these low mass dwarfs 
are extremely faint in these distant clusters, only the next-generation telescopes such 
as the {\sl James Webb Space Telescope} or the {\sl projected Extremely Large Telescope} 
can obtain a high signal to noise ratio measurements.}

\acknowledgements 
{We thank the anonymous reviewer for their helpful comments and suggestions. 
C. L. acknowledges support from the one-hundred-talent project of Sun Yat-Sen 
University. C. L. and Y. W. were supported by the National Natural Science Foundation 
of China under grants 11803048. Y. W. acknowledges the support by China Postdoctoral 
Science Foundation, and the Young Researcher Grant of National Astronomical 
Observatories, Chinese Academy of Sciences.}


\facilities{{\sl Hubble Space Telescope} (UVIS/WFC3 and ACS/WFC)}
\software{\sc dolphot2.0 \citep{Dolp11a,Dolp11b,Dolp13a}}
\\{iSpec \citep{Blan14a,Blan19a}}


\begin{thebibliography}{}
\expandafter\ifx\csname natexlab\endcsname\relax\def\natexlab#1{#1}\fi
\providecommand{\url}[1]{\href{#1}{#1}}
 \bibitem[Barker \& Paust(2018)]{Bark18a} Barker, H., \& Paust,
  N.~E.~Q.\ 2018, \pasp, 130, 034204 

\bibitem[Bastian \& Strader(2014)]{Bast14a} Bastian, N., \& Strader, J.\ 2014, \mnras, 443, 3594 

\bibitem[Bastian et al.(2015)]{Bast15a} Bastian, N., Cabrera-Ziri, I., \& Salaris, M.\ 2015, \mnras, 449, 3333 

\bibitem[Bastian \& Lardo(2018)]{Bast18a} Bastian, N., \& Lardo, C.\ 2018, \araa, 56, 83 

\bibitem[Baumgardt et al.(2013)]{Baum13a} Baumgardt, H., Parmentier,
  G., Anders, P., \& Grebel, E.~K.\ 2013, \mnras, 430, 676


\bibitem[Bekki(2019)]{Bekk19a} Bekki, K.\ 2019, \aap, 622, A53 

\bibitem[Bell et al.(1979)]{Bell79a} Bell, R.~A., Dickens, R.~J., \& Gustafsson, B.\ 1979, \apj, 229, 604 

\bibitem[Blanco-Cuaresma et al.(2014)]{Blan14a} Blanco-Cuaresma, S., Soubiran, C., Heiter, U., \& Jofr{\'e}, P.\ 2014, \aap, 569, A111 

\bibitem[Blanco-Cuaresma(2019)]{Blan19a} Blanco-Cuaresma, S.\ 2019, \mnras, 486, 2075 

\bibitem[Bragaglia et al.(2017)]{Brag17a} Bragaglia, A., Carretta, E., D'Orazi, V., et al.\ 2017, \aap, 607, A44 

\bibitem[Briley et al.(1989)]{Bril89a} Briley, M.~M., Bell, R.~A., Smith, G.~H., \& Hesser, J.~E.\ 1989, \apj, 341, 800 

\bibitem[Cannon et al.(1998)]{Cann98a} Cannon, R.~D., Croke, B.~F.~W., Bell, R.~A., Hesser, J.~E., \& Stathakis, R.~A.\ 1998, \mnras, 298, 601 

\bibitem[Carretta et al.(2009)]{Carr09a} Carretta, E., Bragaglia, A., Gratton, R.~G., et al.\ 2009, \aap, 505, 117 

\bibitem[Carretta et al.(2010)]{Carr10a} Carretta, E., Bragaglia, A., Gratton, R.~G., et al.\ 2010, \aap, 516, A55 

\bibitem[Cavallo et al.(1996)]{Cava96a} Cavallo, R.~M., Sweigart, A.~V., \& Bell, R.~A.\ 1996, \apjl, 464, L79 

\bibitem[Cavallo et al.(1998)]{Cava98a} Cavallo, R.~M., Sweigart, A.~V., \& Bell, R.~A.\ 1998, \apj, 492, 575 

\bibitem[Chantereau et al.(2019)]{Chan19a} Chantereau, W., Salaris, M., Bastian, N., \& Martocchia, S.\ 2019, \mnras, 484, 5236 

\bibitem[Choi et al.(2016)]{Choi16a} Choi, J., Dotter, A., Conroy, C., et al.\ 2016, \apj, 823, 102
  
\bibitem[Dotter(2016)]{Dott16a} Dotter, A.\ 2016, \apjs, 222, 8

\bibitem[Cordoni et al.(2018)]{Cord18a} Cordoni, G., Milone, A.~P., Marino, A.~F., et al.\ 2018, \apj, 869, 139 

\bibitem[Dalessandro et al.(2016)]{Dale16a} Dalessandro, E., Lapenna, E., Mucciarelli, A., et al.\ 2016, \apj, 829, 77 

\bibitem[D'Antona et al.(2017)]{Dant17a} D'Antona, F., Milone, A.~P., Tailo, M., et al.\ 2017, Nature Astronomy, 1, 0186 

\bibitem[Decressin et al.(2007)]{Decr07a} Decressin, T., Meynet, G., Charbonnel, C., Prantzos, N., \& Ekstr{\"o}m, S.\ 2007, \aap, 464, 1029 

\bibitem[D'Ercole et al.(2008)]{Derc08a} D'Ercole, A., Vesperini, E., D'Antona, F., McMillan, S.~L.~W., \& Recchi, S.\ 2008, \mnras, 391, 825 

\bibitem[de Mink et al.(2009)]{deMi09a} de Mink, S.~E., Pols, O.~R., Langer, N., \& Izzard, R.~G.\ 2009, \aap, 507, L1 

\bibitem[Dolphin.(2011a)]{Dolp11a} Dolphin A., DOLPHOT/WFC3 user's
  guide, version 2.0. {\url
    {http://americano.dolphinsim.com/dolphin/dolphotWFC3.pdf}}

\bibitem[Dolphin.(2011b)]{Dolp11b} Dolphin A., DOLPHOT/WFPC2 user's
  guide, version 2.0. {\url
    {http://americano.dolphinsim.com/dolphot/dolphotWFPC2.pdf}}

\bibitem[Dolphin.(2013)]{Dolp13a} Dolphin A., DOLPHOT user's guide,
  version 2.0. {\url
    {http://americano.dolphinsim.com/dolphot/dolphot.pdf}}

\bibitem[Dotter et al.(2018)]{Dott18a} Dotter, A., Milone, A.~P., Conroy, C., Marino, A.~F., \& Sarajedini, A.\ 2018, \apjl, 865, L10 
    
\bibitem[Dupree et al.(2017)]{Dupr17a} Dupree, A.~K., Dotter, A., Johnson, C.~I., et al.\ 2017, \apjl, 846, L1 

\bibitem[For \& Bekki(2017)]{For17a} For, B.-Q., \& Bekki, K.\ 2017, \mnras, 468, L11 

\bibitem[Georgy et al.(2019)]{Geor19a} Georgy, C., Charbonnel, C., Amard, L., et al.\ 2019, \aap, 622, A66 



\bibitem[Gray \& Corbally(1994)]{Gray94a} Gray, R.~O., \& Corbally, C.~J.\ 1994, \aj, 107, 742 

\bibitem[Gustafsson et al.(2008)]{Gust08a} Gustafsson, B., Edvardsson, B., Eriksson, K., et al.\ 2008, \aap, 486, 951 

\bibitem[Hollyhead et al.(2019)]{Holl19a} Hollyhead, K., Martocchia, S., Lardo, C., et al.\ 2019, \mnras, 484, 4718 

\bibitem[Hong et al.(2017)]{Hong17a} Hong, J., de Grijs, R., Askar, A., et al.\ 2017, \mnras, 472, 67 

\bibitem[Jiang et al.(2014)]{Jiang14a} Jiang, D., Han, Z., \& Li, L.\ 2014, \apj, 789, 88 

\bibitem[Krause et al.(2016)]{Krau16a} Krause, M.~G.~H., Charbonnel,
  C., Bastian, N., \& Diehl, R.\ 2016, \aap, 587, A53

\bibitem[Lagioia et al.(2019)]{Lagi19a} Lagioia, E.~P., Milone, A.~P., Marino, A.~F., \& Dotter, A.\ 2019, \apj, 871, 140 

\bibitem[Larsen et al.(2014)]{Lars14a} Larsen, S.~S., Brodie, J.~P., Grundahl, F., \& Strader, J.\ 2014, \apj, 797, 15 


\bibitem[Letarte et al.(2006)]{Leta06a} Letarte, B., Hill, V., Jablonka, P., et al.\ 2006, \aap, 453, 547 

\bibitem[Li et al.(2016)]{Li16a} Li, C., de Grijs, R., Deng, L., et al.\ 2016, \nat, 529, 502 

\bibitem[Li et al.(2017)]{Li17a} Li, C., de Grijs, R., Deng, L., \& Milone, A.~P.\ 2017, \apj, 844, 119 

\bibitem[Li \& de Grijs(2019)]{Li19a} Li, C., \& de Grijs, R.\ 2019, \apj, 876, 94 

\bibitem[Marino et al.(2008)]{Mari08a} Marino, A.~F., Villanova, S., Piotto, G., et al.\ 2008, \aap, 490, 625 

\bibitem[Marino et al.(2009)]{Mari09a} Marino, A.~F., Milone, A.~P., Piotto, G., et al.\ 2009, \aap, 505, 1099 

\bibitem[Marino et al.(2018a)]{Mari18a} Marino, A.~F., Przybilla, N., Milone, A.~P., et al.\ 2018, \aj, 156, 116 

\bibitem[Marino et al.(2018b)]{Mari18b} Marino, A.~F., Milone, A.~P., Casagrande, L., et al.\ 2018, \apjl, 863, L33 

\bibitem[Martocchia et al.(2017)]{Mart17a} Martocchia, S., Bastian, N., Usher, C., et al.\ 2017, \mnras, 468, 3150 

\bibitem[Martocchia et al.(2018)]{Mart18a} Martocchia, S., Cabrera-Ziri, I., Lardo, C., et al.\ 2018, \mnras, 473, 2688 

\bibitem[Martocchia et al.(2019)]{Mart19a} Martocchia, S., Dalessandro, E., Lardo, C., et al.\ 2019, \mnras

\bibitem[McLaughlin \& van der Marel(2005)]{Mcla05a} McLaughlin, D.~E., \& van der Marel, R.~P.\ 2005, \apjs, 161, 304

\bibitem[Milone et al.(2009)]{Milo09a} Milone, A.~P., Bedin, L.~R., Piotto, G., \& Anderson, J.\ 2009, \aap, 497, 755 



\bibitem[Milone et al.(2012)]{Milo12a} Milone, A.~P., Piotto, G., Bedin, L.~R., et al.\ 2012, \aap, 540, A16 

\bibitem[Milone et al.(2017)]{Milo17a} Milone, A.~P., Piotto, G., Renzini, A., et al.\ 2017, \mnras, 464, 3636 

\bibitem[Milone et al.(2018)]{Milo18a} Milone, A.~P., Marino, A.~F., Renzini, A., et al.\ 2018, \mnras, 481, 5098 

\bibitem[Monelli et al.(2013)]{Mone13a} Monelli, M., Milone, A.~P.,
  Stetson, P.~B., et al.\ 2013, \mnras, 431, 2126
  
\bibitem[Mucciarelli et al.(2009)]{Mucc09a} Mucciarelli, A., Origlia, L., Ferraro, F.~R., \& Pancino, E.\ 2009, \apjl, 695, L134 

\bibitem[Mucciarelli et al.(2014)]{Mucc14a} Mucciarelli, A., Dalessandro, E., Ferraro, F.~R., Origlia, L., \& Lanzoni, B.\ 2014, \apjl, 793, L6 

\bibitem[Niederhofer et al.(2017)]{Nied17a} Niederhofer, F., Bastian, N., Kozhurina-Platais, V., et al.\ 2017, \mnras, 465, 4159 

\bibitem[Pancino et al.(2017)]{Panc17a} Pancino, E., Romano, D., Tang, B., et al.\ 2017, \aap, 601, A112 

\bibitem[Paxton et al.(2011)]{Paxt11a} Paxton, B., Bildsten, L.,
  Dotter, A., et al.\ 2011, \apjs, 192, 3

\bibitem[Paxton et al.(2013)]{Paxt13a} Paxton, B., Cantiello, M.,
  Arras, P., et al.\ 2013, \apjs, 208, 4

\bibitem[Paxton et al.(2015)]{Paxt15a} Paxton, B., Marchant, P.,
  Schwab, J., et al.\ 2015, \apjs, 220, 15

\bibitem[Piatti(2018)]{Piat18a} Piatti, A.~E.\ 2018, \aj, 156, 206 

\bibitem[Pilachowski et al.(1996)]{Pila96a} Pilachowski, C.~A., Sneden, C., Kraft, R.~P., \& Langer, G.~E.\ 1996, \aj, 112, 545 

\bibitem[Piotto et al.(2015)]{Piot15a} Piotto, G., Milone, A.~P., Bedin, L.~R., et al.\ 2015, \aj, 149, 91 

\bibitem[Piotto et al.(2007)]{Piot07a} Piotto, G., Bedin, L.~R., Anderson, J., et al.\ 2007, \apjl, 661, L53

\bibitem[Sbordone et al.(2007)]{Sbor07a} Sbordone, L., Bonifacio, P., Buonanno, R., et al.\ 2007, \aap, 465, 815 

\bibitem[Smith et al.(2006)]{Smit06a} Smith, L.~J., Westmoquette, M.~S., Gallagher, J.~S., et al.\ 2006, \mnras, 370, 513 

\bibitem[Sun et al.(2019)]{Sun19a} Sun, W., de Grijs, R., Deng, L., \& Albrow, M.~D.\ 2019, \apj, 876, 113 

\bibitem[Sweigart \& Mengel(1979)]{Swei79a} Sweigart, A.~V., \& Mengel, J.~G.\ 1979, \apj, 229, 624 

\bibitem[Tautvai{\v s}ien{\.e} et al.(2004)]{Taut04a} Tautvai{\v s}ien{\.e}, G., Wallerstein, G., Geisler, D., Gonzalez, G., \& Charbonnel, C.\ 2004, \aj, 127, 373 

\bibitem[Villanova et al.(2007)]{Vill07a} Villanova, S., Piotto, G., King, I.~R., et al.\ 2007, \apj, 663, 296

\bibitem[Villanova et al.(2013)]{Vill13a} Villanova, S., Geisler, D., Carraro, G., Moni Bidin, C., \& Mu{\~n}oz, C.\ 2013, \apj, 778, 186   

\bibitem[Wang et al.(2017)]{Wang17a} Wang, Y., Primas, F., Charbonnel, C., et al.\ 2017, \aap, 607, A135 

\bibitem[Yang et al.(2013)]{Yang13a} Yang, W., Bi, S., Meng, X., \& Liu, Z.\ 2013, \apj, 776, 112 

\bibitem[Zhang et al.(2018)]{Zhang18a} Zhang, H., de Grijs, R., Li, C., \& Wu, X.\ 2018, \apj, 853, 186 

\end{thebibliography}
\end{document}